\documentclass[12pt, draftclsnofoot, onecolumn]{IEEEtran}
\usepackage{amsmath,amssymb}
\usepackage{amsfonts}
\usepackage{amsbsy}
\usepackage{graphicx}
\usepackage{caption}

\usepackage{subfigure}
\usepackage{stfloats}
\usepackage{epstopdf}
\usepackage{color}
\usepackage{etoolbox}
\makeatletter
\patchcmd{\@maketitle}
{\addvspace{0.5\baselineskip}\egroup}
{\addvspace{-1\baselineskip}\egroup}
{}
{}
\makeatother
\begin{document}
	
	\title{\huge Outage Performance of Multi-UAV Relaying-based Imperfect Hardware Hybrid Satellite-Terrestrial Networks}
	
	\author{\IEEEauthorblockN{Pankaj K. Sharma and Deepika Gupta} 
		\thanks{Pankaj K. Sharma is with the Department of Electronics and Communication Engineering, National Institute of Technology Rourkela, India. Email: sharmap@nitrkl.ac.in.}
		\thanks{Deepika Gupta is with the Department of Electronics and Communication Engineering, Dr S P M International Institute of Information Technology, Naya Raipur, India. Email: deepika@iiitnr.edu.in}
		
	}
	\maketitle
	
	\begin{abstract}
		In this paper, we consider an imperfect hardware hybrid satellite-terrestrial network (HSTN) where the satellite communication with a ground user equipment (UE) is aided by the multiple amplify-and-forward (AF) three-dimensional ($3$D) mobile unmanned aerial vehicle (UAV) relays. Herein, we consider that all transceiver nodes are corrupted by the radio frequency hardware impairments (RFHI). Further, a stochastic mixed mobility (MM) model is employed to characterize the instantaneous location of $3$D mobile UAV relays in a cylindrical cell with UE lying at its center on ground plane. Taking into account the aggregate RFHI model for satellite and UAV relay transceivers and the random $3$D distances-based path loss for UAV relay-UE links, we investigate the outage probability (OP) and corresponding asymptotic outage behaviour of the system under an opportunistic relay selection scheme in a unified form for shadowed-Rician satellite links' channels and Nakagami-\emph{m} as well as Rician terrestrial links' channels. We corroborate theoretical analysis by simulations.
		
	\end{abstract}
	
	\section{Introduction}
	\IEEEPARstart{H}{ybrid} satellite-terrestrial networks (HSTNs) taking into account the terrestrial relaying to aid satellite communication have received great importance for mitigating the deleterious masking effect \cite{r9}. 
	Therefore, the works \cite{bhatna}-\cite{pks1} have recently analyzed the performance of static relay-based HSTNs with perfect transceiver hardware. However, the transceivers (e.g., low-complexity relays) with imperfect hardware subject to radio-frequency hardware impairments (RFHI) \cite{eb} resulting from IQ imbalances, amplifier non-linearities, phase noise, etc.
	Consequently, the performance of terrestrial networks with RFHI and HSTNs was analyzed in works \cite{pan5} and \cite{guo}, respectively. 
	On another hand, mobile unmanned aerial vehicle (UAV) relaying has recently been considered as a key candidate for future wireless communications \cite{yzen}. Due to low cost, flexible deployment, and line-of-sight (LOS) communications, the small rotary-wing UAVs have qualified as potential candidates for mobile relaying. Eventually, the mobile UAV relaying has been recently employed in \cite{r13} for terrestrial networks and in \cite{r12}, \cite{r11} for HSTNs. Note that along with the satellite, the UAVs may subject to RFHI due to limited onboard sophisticated signal processing equipment. However, the impact of RFHI on UAV relaying-based HSTNs has not been investigated in literature.  
	
	In contrast to previous decode-and-forward relaying-based HSTN in \cite{r11}, this paper investigates the outage probability (OP) and corresponding asymptotic outage behaviour of an \textit{imperfect hardware HSTN system model with low-complexity amplify-and-forward (AF) mobile UAV relays}. Herein, we resort to the aggregate RFHI model proposed in \cite{eb} where the  satellite transmitter and UAV receiver distortions are lumped together at satellite node. Likewise, the UAV transmitter and ground user equipment (UE) distortions are lumped together at UAV node. We adopt the stochastic mixed mobility (MM) from \cite{pank} to station UAV relays in a cylindrical space at a given snapshot of time for calculating the path loss based on three-dimensional ($3$D) UAV-to-ground user distance distribution. \textit{More importantly, we present our analysis in a unified form for shadowed-Rician satellite links' and Nakagami-\emph{m} as well as analytically challenging Rician terrestrial links' fading channels which was missing in \cite{r11}. Although the unified OP expression is exact only for terrestrial Nakagami-\emph{m} fading, it is an accurate finite-series solution for otherwise cumbersome terrestrial Rician fading.} We present useful insights through our analysis on the outage performance of considered HSTN.
	\section{System Description}\label{sysmod}
	
	We consider an HSTN where information transmission from a geostationary earth orbit (GEO) satellite $S$ to a fixed-location ground user equipment (UE) $D$ is facilitated by one of the opportunistically selected UAV relays $U_i$, among $M$ candidate relays $i=1,...,M$ deployed at $3$D locations in a cylindrical cell of height $H$ and radius $R$. The direct link from $S$ to $D$ is assumed to be heavily masked. Further, we assume that the satellite transmitter $S$, UAV relay transceivers $U_i$, and UE $D$ are built using simple hardware equipment resulting in RFHI. As followed in \cite{eb}, instead of considering individual RFHI components at various transmitter/receiver nodes, we consider the lumped distortions at $S$ for satellite transmitter and UAV receiver distortions, and at $U_i$ the corresponding UAV transmitter and UE receiver distortions. Further, the $3$D cylindrical cell lies within the footprint of $S$ around $D$ which is placed at its centre in ground plane. We station the UAV relays randomly in the $3$D deployment region by first running the MM model \cite{pank} for sufficient time duration, and then, taking a snapshot at specific observation time. The channels from $S$ to $U_{i}$ and $U_i$ to $D$ are denoted by $g_{su_i}$  and ${g}_{u_id}$, respectively. The noise at various nodes is additive white Gaussian noise (AWGN) with zero mean and variance $\sigma^2$.    
	\subsection{Mixed Mobility Model for UAV}
	We consider the multi-parameter MM model \cite{pank} which describes the $3$D mobility of UAVs in a cylinder of height $H$ and radius $R$. The UAV makes vertical transitions based on random waypoint mobility (RWPM) model with random dwell time at each waypoint. at time $t$, the pdf of instantaneous altitude of UAV $h_{i}(t)$ is given by the weighted sum of a static pdf $f^{st}_{h_i}(x|t)$ and a mobility pdf $f^{mo}_{h_i}(x|t)$ as
	$f_{h_{i}}(x|t)=p_\mathrm{s}f^{st}_{h_i}(x|t)+(1-p_\mathrm{s})f^{mo}_{h_i}(x|t)$,
	where 
	$f^{st}_{h_{i}}(x|t)=\frac{1}{H}$
	and
	$f^{mo}_{h_{i}}(x|t)=-\frac{6x^2}{H^3}+\frac{6x}{H^2}$, for $0 \leq x \leq H,$
	with weight $p_\mathrm{s}=\frac{\mathbb{E}[T_\mathrm{s}]}{\mathbb{E}[T_\mathrm{s}]+\mathbb{E}[T_\mathrm{m}]}$ as stay probability at waypoints. 
	
	Meanwhile, in the dwell time, the UAV makes random walk (RW) in horizontal plane by following $z_i(t+1)=z_{i}(t)+\mathrm{u}_i(t)$ with probability $p_\mathrm{s}$,
	where $z_i(t)$ denotes the projection of UAV's location on ground plane and $\mathrm{u}_i(t)$ is the uniform distribution in ball $B(z_i(t),R^{\prime})$ with
	$R^\prime$ as the maximum spatial mobility range. Whereas, it follows $z_i(t+1)=z_{i}(t)$ with probability $1-p_\mathrm{s}$. Consequently, the pdf of distance $Z_i(t)=\|z_i(t)\|$ is given by 
	$f_{Z_i}(z|t)=\frac{2z}{R^2}, 0\leq z \leq R$.
	Various parameters associated with the MM model are described below:
	$v_{1,i}(t)\sim[v_{min},v_{max}]$: Uniformly random velocity of vertical transition at waypoints; 	$T_\mathrm{s}\sim[\tau_{min}, \tau_{max}]$ and $\mathbb{E}[T_\mathrm{s}]$: Uniformly random and mean dwell time; $T_\mathrm{m}$ and $\mathbb{E}[T_\mathrm{m}]=\frac{\ln(v_{max}/v_{min})}{v_{max}-v_{min}}\,\frac{H}{3}$: Random and mean vertical movement time; and ${v}_{2,i}(t)=\|z_i(t)-z_{i}(t-1)\|$ and $\mathbb{E}[{v}_{2,i}(t)]=\frac{R^\prime}{1.5}$: Random and mean velocity of horizontal transition at waypoints.  
	\subsection{Channel Models}
	\subsubsection{Satellite-UAV links}
	For $g_{su_{i}}$ following SR fading, the probability density function (pdf) of $|g_{su_{i}}|^{2}$ is given by
	\begin{align}\label{pdfrsaa}
		f_{|g_{su_{i}}|^{2}}(x)=\alpha_{u} \sum_{\kappa=0}^{m_{su}-1}\zeta(\kappa)x^{\kappa}\textmd{e}^{-(\beta_{u}-\delta_{u})x},
	\end{align}
	where $\alpha_{u}=(2\flat_{su} m_{su}/(2\flat_{su} m_{su}+\Omega_{su}))^{m_{su}}/2\flat_{su} $, $\beta_{u}=1/2\flat_{su} $, and $\delta_{u}=\Omega_{su}/(2\flat_{su} )(2\flat_{su} m_{su}+\Omega_{su})$, $\Omega_{su}$ and $2\flat_{su} $ are the average powers, respectively, of the LOS and multipath components, $m_{su}$ is the fading severity,  $\zeta(\kappa)=(-1)^{\kappa}(1-m_{su})_{\kappa}\delta_{u}^{\kappa}/(\kappa!)^{2}$, and $(\cdot)_{\kappa}$ denotes the Pochhammer symbol \cite[p. xliii]{grad}.
	Additionally, a free space loss scale factor for satellite links is given as \cite{guo} 
	$\sqrt{\mathcal{L}_{su_i}(t)\vartheta_s\vartheta(\theta_{u_i})}=\sqrt{\frac{\vartheta_s\vartheta(\theta_{u_i})}{\mathcal{K_{B}}\mathcal{TW}}}\left(\frac{c}{4\pi f_\mathrm{c} \mathrm{d}_{u_i}(t)}\right)$,
	where $\mathcal{K_B}=1.38\times10^{-23}$J/K is the Boltzman constant, $\mathcal{T}$ is the receiver noise temperature, $\mathcal{W}$ is the carrier bandwidth, $c$ is the speed of light, $f_\mathrm{c}$ is the carrier frequency, and $\mathrm{d}_{u_i}(t)$ is the distance between $S$ and $U_i$. Here, $\vartheta_s$ denotes the antenna gain at satellite, $\vartheta(\theta_{u_i})$ gives the beam gain of satellite towards $U_i$ which can be expressed as
	$\vartheta(\theta_{u_i})=\vartheta_{u_i}\left(\frac{\mathcal{J}_1(\rho_{u_i})}{2\rho_{u_i}}+36\frac{\mathcal{J}_3(\rho_{u_i})}{\rho^3_{u_i}}\right)$,
	where $\theta_{u_i}$ is the angular separation of $U_i$ from the satellite beam center, $\vartheta_{u_i}$ is the antenna gain at $U_i$, $\mathcal{J}_\varrho(\cdot)$, $\varrho\in\{1,3\}$ is the Bessel function, and $\rho_{u_i}=2.07123\frac{\sin \theta_{u_i}}{\sin \theta_{{u_i}3\textmd{dB}}}$ with $\theta_{{u_i}3\text{dB}}$ as $3$dB beamwidth.
	
	\subsubsection{UAV-Ground UE links}
	The Nakagami-\emph{m} fading for terrestrial $U_i$ to $D$ links yields the pdf of $|g_{u_{i}d}|^2$ as 
	\begin{align}\label{gam}
		f_{|g_{u_{i}d}|^2}(x)&=\left( \frac{m_{ud}}{\Omega_{ud}}\right)^{m_{ud}} \frac{x^{m_{ud}-1}}{\Gamma(m_{ud})}\,\textmd{e}^{ -\frac{m_{ud}}{\Omega_{ud}} x} ,
	\end{align}
	where $m_{ud}$ and $\Omega_{ud}$ represent the fading severity and average channel power, respectively. Further, the Rician fading of $U_i$ to $D$ links results in the pdf of $|g_{u_{i}d}|^2$ as  
	\begin{align}\label{gamq}
		f_{|g_{u_{i}d}|^2}(x)&=\frac{(1+K_{ud})}{\Omega_{ud}}\textmd{e}^{-K_{ud}}\textmd{e}^{-\frac{(1+K_{ud})x}{\Omega_{ud}}}\\\nonumber
		&\times\mathcal{I}_0\left(2\sqrt{{K_{ud}(1+K_{ud})x}/{\Omega_{ud}}}\right),
	\end{align}
	where $K_{ud}$ is the Rician $K$-factor and $\mathcal{I}_0(\cdot)$ is the zeroth-order modified Bessel function of first kind \cite[eq. 9.6.10]{grad}. Next, the instantaneous free-space path loss from $U_i$ to $D$ can be given as
	$W^{-\alpha}_{id}(t)={\left({h^2_i(t)+Z^2_i(t)}\right)^{-\frac{\alpha}{2}}}$,
	where $w_{id}$ is the distance from $U_i$ to $D$ with $\alpha$ as path loss exponent. Due to space constraint, we consider the path loss model with unity LOS probability in this letter. 
	
	
	\subsection{Propagation Model}
	In considered RFHI-affected HSTN, a satellite $S$ communicates with UE $D$ in two consecutive time slots through a selected AF UAV relay $U_i$. In the first slot at time $t$, $S$ sends the message $x_{s}(t)$ with power $P_s=\mathbb{E}[|x^2_{s}(t)|]$ to $U_i$. The received message signal at $U_i$ under lumped RFHI distortions at $S$ can be given as
	\begin{align}\label{3}
		y_{u_i}(t)\!&=\sqrt{\mathcal{L}_{su_i}\!(t)\vartheta_s\vartheta(\theta_{u_i})}{g}_{su_i}(t)(x_{s}(t)+\rho_{{s_i}})+\nu_{u_i},
	\end{align}
	where $\rho_{{s_i}}\sim\mathcal{CN}(0, \kappa^2_{{s_i}}P_s)$ is the aggregate distortion noise at $S$ with $\kappa_{{s_i}}$ as RFHI parameter, 
	$\vartheta_s$ is the satellite antenna gain, $\nu_{u_i}$ is the AWGN at $U_i$.  
	
	In the second slot at time $t+1$, the $U_i$ amplifies the received signal $y_{u_i}(t)$ with a gain factor $G$ and forward the amplified signal to UE $D$. Thus, the received signal at $D$ can be expressed as
	\begin{align}\label{6}
		y_{id}(t\!+\!1)&=W^{-\frac{\alpha}{2}}_{id}(t\!+\!1){g}_{u_id}(t\!+\!1)(Gy_{u_i}(t)\!+\!\rho_{u_i})\!+\!\nu_{d},
	\end{align}
	where $\rho_{u_i}\sim\mathcal{CN}(0, \kappa^2_{{u_i}}P_u)$, is the aggregate distortion noise at $U_i$ with $\kappa_{{u_i}}$ as RFHI parameter, $\nu_{d}$ is the AWGN at $D$. Here, the gain factor $G$ can be computed as $G=\sqrt{\frac{P_u}{P_s\mathcal{L}_{su_i}\!(t)\vartheta_s\vartheta(\theta_{u_i})|g_{s_{u_i}}|^2(1+\kappa^2_{s_i})+\sigma^2}}$ to calculate the end-to-end signal-to-noise plus distortion ratio (SNDR) using (\ref{6}) as
	\begin{align}\label{7}
		&\Lambda_{id}(t+1)=\\\nonumber
		&\frac{\Lambda_{su_i}(t)\Lambda_{u_id}(t+1)}{\lambda_1\Lambda_{su_i}(t)\Lambda_{u_id}(t\!+\!1)\!+\!\lambda_2\Lambda_{su_i}(t)\!+\!\lambda_3\Lambda_{u_id}(t\!+\!1)\!+\!1},
	\end{align}
	where $\Lambda_{su_i}(t)=\frac{P_s\mathcal{L}_{su_i}(t)\vartheta_s\vartheta(\theta_{u_i})|{g}_{su_i}(t)|^2}{\sigma^2}$, $\Lambda_{u_id}(t+1)=\frac{P_u W^{-\alpha}_{id}(t+1)|{g}_{u_id}(t+1)|^2}{\sigma^2}$, $\lambda_1=\kappa^2_{s_i}+\kappa^2_{u_i}+\kappa^2_{s_i}\kappa^2_{u_i}$, $\lambda_2=1+\kappa^2_{s_i}$, and $\lambda_3=1+\kappa^2_{u_i}$. For analytical tractability, as in \cite{ttd}, we consider RFHI levels $\kappa_{s_i}=\kappa_{s}$ and $\kappa_{u_i}=\kappa_{u}$ $\forall i$. Further, we observe that the distance $\mathrm{d}_{u_i}(t)$ is very large (e.g., $35,786$ Km, for GEO satellite), it can be reasonably considered $\mathrm{d}_{u_i}(t)\approx\mathrm{d}_{u}(t)$, $\mathcal{L}_{su_i}(t)\approx\mathcal{L}_{su}(t)$, $\theta_{u_i}\approx\theta_{u}$, $\rho_{u_i}\approx\rho_{u}$, and $\vartheta_{u_i}\approx\vartheta_{u}$, $\forall i$ to carry-out the OP analysis under independent and identically distributed (i.i.d.) channels.
	
	
	Let an opportunistic UAV relay selection be applied to maximize the SNDR at $D$ as $i^\star(t)=\displaystyle \arg\max_{i\in1,...,M} \Lambda_{id}(t)$.
	
	\section{Outage Probability Analysis}\label{eop}
	We next drop the time notation $t$ to analyze the OP in one snapshot under causal $S$ to $D$ transmission (i.e., $t>1$).
	
	We first present the important cumulative distribution function (cdf) and pdf expressions to be used later in this paper.
	
	The cdf $F_{\Lambda_{su_i}}(x)$ can be expressed, by applying a variable transformation for $\Lambda_{su_i}=\eta_{u}|g_{su_i}|^{2}$ in (\ref{pdfrsaa}), as  
	\begin{align}\label{pdflsz}
		F_{\Lambda_{su_i}}(x)&=1-\alpha_{u}\!\!\!\!\sum_{\kappa=0}^{m_{su}-1}\!\!\!\!\frac{\zeta(\kappa)}{(\eta_{s})^{\kappa+1}}\sum_{p=0}^{\kappa} \frac{\kappa!}{p!}{\Theta_{u}}^{-(\kappa+1-p)}\\\nonumber
		&\times x^{p}\textmd{e}^{-\Theta_{u}x},
	\end{align}
	where $\Theta_{u}=\frac{\beta_{u}-\delta_{u}}{\eta_{s}}$ and  $\eta_{s}=$ $\frac{P_s\mathcal{L}_{su}\vartheta_s\vartheta(\theta_{u})}{\sigma^2}$.  
	
	Furthermore, the pdf of random distance $W_{id}$ under MM model \cite{pank} is given as 
	\begin{align}\label{1u9}
		f_{W_{id}}(w)&=p_{s}f^{st}_{W_{id}}(w)+(1-p_{s})f^{mo}_{W_{id}}(w),
	\end{align}
	where the pdfs $f^{st}_{W_{id}}(w)$ and $f^{mo}_{W_{id}}(w)$ correspond to the UAV's horizontal and vertical motions, respectively, and are given as  
	\begin{align}\label{ju11}
		f^{st}_{W_{id}}(w)&=\left\{ \begin{array}{l}
			\frac{2w^{2}}{R^2 H},\textmd{ for }0\leq w<H,\\
			\frac{2w}{R^2},\textmd{ for }H\leq w<R,\\
			\frac{2w}{R^2}-\frac{2w\sqrt{w^2-R^2}}{R^2H},\\
			\qquad\textmd{ for } R\leq w\leq\sqrt{R^2+H^2}.
		\end{array}\right.
	\end{align}
	\begin{align}\label{thuo11}
		\textmd{and }f^{mo}_{W_{id}}(w)&=\left\{ \begin{array}{l}
			\frac{6w^{3}}{R^2 H^2}-\frac{4w^{4}}{R^2 H^3},\textmd{ for }0\leq w<H,\\
			\frac{2w}{R^2},\textmd{ for }H\leq w<R,\\
			\frac{2w}{R^2}-\frac{6w(w^2-R^2)}{R^2 H^2}+\frac{4w(w^2-R^2)^{\frac{3}{2}}}{R^2 H^3},\\
			\qquad\textmd{ for } R\leq w\leq\sqrt{R^2+H^2}.
		\end{array}\right.
	\end{align}
	
	In addition, letting $\eta_u=\frac{P_u}{\sigma^2}$, the pdf $f_{\Lambda_{u_{i}d}}(x)$ can be derived using $f_{\Lambda_{u_{i}d}}(x)= \frac{d}{dx}F_{\Lambda_{u_{i}d}}(x)$, where $F_{\Lambda_{u_{i}d}}(x)=\textmd{Pr}\left[{\eta_u W^{-\alpha}_{{i}d}|g_{u_{i}d}|^2}<x\right]$ for Nakagami-\emph{m} and Rician fading can be calculated based on the pdfs given in (\ref{gam}) and (\ref{gamq}) as
	\begin{align}\label{cdfh}
		F_{\Lambda_{u_{i}d}}(x)	&=\int_{0}^{\sqrt{R^2+H^2}}\frac{\Upsilon\left(m_{ud}, \frac{m_{ud}xr^{\alpha}}{\Omega_{ud}\eta_u}\right)}{\Gamma(m_{ud})} f_{W_{id}}(r)dr
	\end{align}
	\begin{align}\label{cdfq}
		&\textmd{and }F_{\Lambda_{u_{i}d}}(x)=\\\nonumber
		&1-\int_{0}^{\sqrt{R^2+H^2}}\mathcal{Q}_1\left(\sqrt{2K_{ud}},\sqrt{\frac{2(1+K_{ud})xr^{\alpha}}{\Omega_{ud}\eta_u}}\right) f_{W_{id}}(r)dr,
	\end{align}
	respectively, where $\mathcal{Q}_1(\cdot,\cdot)$ is the first-order Marcum-Q function \cite{r13}. Using (\ref{cdfh}), we get the pdf
	\begin{align}\label{pdf1}
		f_{\Lambda_{u_{i}d}}(x)&=\frac{1}{\Gamma(m_{ud})}\left(\frac{m_{ud}}{\Omega_{ud}\eta_u}\right)^{m_{ud}}x^{m_{ud}-1}\\\nonumber
		&\times\int_{0}^{\sqrt{R^2+H^2}}r^{m_{ud}\alpha}\textmd{e}^{-\frac{m_{ud}xr^\alpha}{\Omega_{ud}\eta_u} }\!\!\!f_{W_{id}}(r)dr,
	\end{align}
	for Nakagami-\emph{m} fading. To get the tractable pdf for Rician fading, in (\ref{cdfq}), we first apply the approximation  
	$\mathcal{Q}_1(a,b)
	\approx\sum_{r_1=0}^{R_1}\chi_{r_1}\Gamma\left(1+r_1,\frac{b^2}{2}\right)$  \cite{r13},
	where $\chi_{r_1}=\frac{\Gamma(r_1+R_1)R_{1}^{1-2r_1}a^{2r_1}2^{-r_1}}{\Gamma(r_1+1)^2\Gamma(R_1-r_1+1)\textmd{e}^{\frac{a^2}{2}}}$ and the term $R_1$ can be truncated as $50\max(1,a,b)$. Then, taking the derivative, we get the pdf 	 
	\begin{align}\label{pdf2}
		f_{\Lambda_{u_{i}d}}(x)&\approx\sum_{r_1=0}^{R_1}\chi_{r_1}\left(\frac{1+K_{ud}}{\Omega_{ud}\eta_u}\right)^{r_1+1}x^{r_1}\\\nonumber
		&\times\int_{0}^{\sqrt{R^2+H^2}}r^{(r_1+1)\alpha}\textmd{e}^{-\frac{(1+K_{ud})xr^\alpha }{\Omega_{ud}\eta_u}}\!\!\!f_{W_{id}}(r)dr,
	\end{align}
	where for calculating $R_1$, $r$ can be set as upper integral limit.
	After a careful observation of pdf expressions in (\ref{pdf1}) and (\ref{pdf2}), we represent them in a unified mathematical form as 
	\begin{align}\label{pdf3}
		f_{\Lambda_{u_{i}d}}(x)&\approx\sum_{r_1=\lambda_4}^{\lambda_5}\mathcal{B}_{r_1}\mathcal{A}^{r_1+1}x^{r_1}\int_{0}^{\sqrt{R^2+H^2}}r^{(r_1+1)\alpha}\\\nonumber
		&\times\textmd{e}^{-{\mathcal{A}xr^\alpha }}f_{W_{id}}(r)dr,
	\end{align}
	where $\{\lambda_4,\lambda_5, \mathcal{A}, \mathcal{B}_{r_1}\}=\{m_{ud}-1, m_{ud}-1, \frac{m_{ud}}{\Omega_{ud}\eta_u}, \frac{1}{\Gamma(m_{ud})}\}$ and $\{0, R_1, \frac{1+K_{ud}}{\Omega_{ud}\eta_u}, \chi_{r_1}\}$ for Nakagami-\emph{m} and Rician fading, respectively. Here, the analysis remains exact for Nakagami-\emph{m} fading and tight approximation holds for Rician fading only.
	%
	
	\subsection{Outage Probability}
	For a threshold $\gamma_{\textmd{th}}$ and i.i.d. links, we derive the OP for considered HSTN in a unified form using (\ref{7}), (\ref{pdflsz}), and (\ref{pdf3}) as 
	\begin{align}\label{8}
		&\mathcal{P}_{\textmd{out}}(\gamma_{\textmd{th}})=[\textmd{Pr}\left[\Lambda_{id}<\gamma_{\textmd{th}}\right]]^M\\\nonumber
		&=\!\!\left[1\!-\!\!\int_{0}^{\infty}\!\!\left[1\!-\!F_{\Lambda_{su_{i}}}\!\! \left(\!{\lambda_6\!\left(1\!+\!\frac{\lambda_7}{y}\right)}\!\right)\right]\!\! f_{\Lambda_{u_{i}d}}\!\!\left(\frac{y\!+\!\lambda_8}{\lambda_9}\right)\!\frac{dy}{\lambda_9}\right]^M,
	\end{align}
	where $\{\lambda_6, \lambda_7, \lambda_8,\lambda_9\}$$=$$\{\frac{\lambda_3\gamma_{\textmd{th}}}{1-\lambda_1\gamma_{\textmd{th}}},\lambda_2\gamma_{\textmd{th}}+\frac{{1-\lambda_1\gamma_{\textmd{th}}}}{\lambda_3},\lambda_2\gamma_{\textmd{th}},1-\lambda_1\gamma_{\textmd{th}}\}$. 
	\newtheorem{theorem}{Theorem}
	\begin{theorem}\label{th1}
		For $\gamma_{\textmd{th}}<\frac{1}{\lambda_1}$, the OP in (\ref{8}) can be derived as 
		\begin{align}\label{pouyt1}
			&\mathcal{P}_{\textmd{out}}(\gamma_{\textmd{th}})=[1-2\alpha_u\sum_{\kappa=0}^{m_{su}-1}\frac{\zeta(\kappa)}{(\eta_{s})^{\kappa+1}}\sum_{p=0}^{\kappa} \frac{\kappa!}{p!}\sum_{q=0}^{p}\binom{p}{q}\\\nonumber
			&\times\textmd{e}^{-\frac{\Theta_u\lambda_3\gamma_{\textmd{th}}}{1-\lambda_1\gamma_{\textmd{th}}}}\sum_{r_1=\lambda_4}^{\lambda_5}\mathcal{B}_{r_1}\sum_{n=0}^{r_1}\binom{r_1}{n}\left(\frac{\mathcal{A}}{1-\lambda_1\gamma_{\textmd{th}}}\right)^{r_1-\frac{n-q-1}{2}}\\\nonumber
			&\times\Theta^{\frac{n-q+1}{2}-(\kappa+1-p)}_u\left(\frac{\lambda_3\gamma_{\textmd{th}}}{1-\lambda_1\gamma_{\textmd{th}}}\right)^{p+\frac{n-q+1}{2}}\\\nonumber
			&\times\left(\lambda_2\gamma_{\textmd{th}}+\frac{1-\lambda_1\gamma_{\textmd{th}}}{\lambda_3}\right)^{\frac{n+q+1}{2}}(\lambda_2\gamma_{\textmd{th}})^{r_1-n}\\\nonumber
			&\times\int_{0}^{\sqrt{R^2+H^2}}r^{\left(r_1-\frac{n-q-1}{2}\right)\alpha}\textmd{e}^{-{\mathcal{A}r^\alpha}\left(\frac{\lambda_2\gamma_{\textmd{th}}}{1-\lambda_1\gamma_{\textmd{th}}}\right)}\\\nonumber
			&\times\mathcal{K}_{n-q+1}\left(2\sqrt{\frac{\mathcal{A}\Theta_u\lambda_3\gamma_{\textmd{th}}r^\alpha}{(1-\lambda_1\gamma_{\textmd{th}})^2}\left(\lambda_2\gamma_{\textmd{th}}+\frac{1-\lambda_1\gamma_{\textmd{th}}}{\lambda_3}\right)}\right)\\\nonumber
			&\times\left.f_{W_{id}}(r)dr\right]^M,
		\end{align}
		where $\mathcal{K}_v(\cdot)$ is the $v$th-order Bessel function \cite[eq. 8.44]{grad}. 	
	\end{theorem}
	\begin{IEEEproof}
		For the condition $\gamma_{\textmd{th}}<\frac{1}{\lambda_1}$, utilizing (\ref{pdflsz}) and (\ref{pdf1}) along with the relevant parameters $\lambda_j$, $j\in\{6,7,8,9\}$ in (\ref{8}) followed by a change in the order of integration and applying \cite[eq. 3.471.9]{grad}, we obtain the required OP. 
	\end{IEEEproof}
	
	
	To reveal insights on system diversity order, we simplify the OP in Theorem \ref{th1} at high SNDR ($\eta_s, \eta_u\rightarrow \infty$).
	\newtheorem{corollary}{Corollary}
	\begin{corollary}
		For $\gamma_{\textmd{th}}<\frac{1}{\lambda_1}$, the unified asymptotic OP expression in can be obtained as
		\begin{align}\label{poutas}
			\mathcal{P}_{\textmd{out}}(\gamma_{\textmd{th}})&{\simeq} \left[\frac{\alpha_u}{\eta_s}\left(\!\frac{\lambda_2\gamma_{\textmd{th}}}{1\!-\!\lambda_1\gamma_{\textmd{th}}}\!\right)\!+\!\sum_{r_1=\lambda_4}^{\lambda_5}\frac{\mathcal{B}_{r_1}\mathcal{A}^{r_1+1}}{r_1+1}\right.\\\nonumber
			&\times\left.\left(\!\frac{\lambda_3\gamma_{\textmd{th}}}{1\!-\!\lambda_1\gamma_{\textmd{th}}}\!\right)^{r_1+1}\!\!\!\!\int_{0}^{\sqrt{R^2+H^2}}r^{(r_1+1)\alpha}f_{W_{id}}(r)dr\right]^M,
		\end{align}
		where $\{\lambda_4,\lambda_5, \mathcal{A}, \mathcal{B}_{r_1}\}=\{m_{ud}-1, m_{ud}-1, \frac{m_{ud}}{\Omega_{ud}\eta_u}, \frac{1}{\Gamma(m_{ud})}\}$ and $\{0, 0, \frac{1+K_{ud}}{\Omega_{ud}\eta_u}, \chi_{r_1}\}$ for Nakagami-\emph{m} and Rician fading, respectively. 
	\end{corollary}
	\begin{IEEEproof}
		We first apply the bound $\frac{xy}{x+y}< \min (x,y)$ (after neglecting $1$ from denominator) in (\ref{7}) to get $\Lambda_{id}\leq\frac{1}{\lambda_1+{1}/{\min\left({\Lambda_{su_i}}/{\lambda_2},{\Lambda_{u_id}}/{\lambda_3}\right)}}$. Then, invoking this in $\mathcal{P}_{\textmd{out}}(\gamma_{\textmd{th}})=[\textmd{Pr}\left[\Lambda_{id}<\gamma_{\textmd{th}}\right]]^M$,  for $\gamma_{\textmd{th}}<\frac{1}{\lambda_1}$, we get
		$\mathcal{P}_{\textmd{out}}(\gamma_{\textmd{th}})
		{\simeq}\! \left[F_{\Lambda_{su_i}}\!\!\left(\!\frac{\lambda_2\gamma_{\textmd{th}}}{1\!-\!\lambda_1\gamma_{\textmd{th}}}\!\right)\!+\!F_{\Lambda_{u_id}}\!\!\left(\!\frac{\lambda_3\gamma_{\textmd{th}}}{1\!-\!\lambda_1\gamma_{\textmd{th}}}\!\right)\right]^M,$
		where the product of cdfs is neglected. Herein, for small $x$, we simplify the cdf  $F_{\Lambda_{su_i}}(x)\simeq\frac{\alpha_u x}{\eta_s}$ \cite{pks}. Further, we have $F_{\Lambda_{u_id}}(x)\simeq\sum_{r_1=\lambda_4}^{\lambda_5}\frac{\mathcal{B}_{r_1}\mathcal{A}^{r_1+1}}{r_1+1}x^{r_1+1}\int_{0}^{\sqrt{R^2+H^2}}r^{(r_1+1)\alpha}f_{W_{id}}(r)dr$ using the pdf in (\ref{pdf3}), for small $x$. Furthermore, for Rician fading, the simplification for small $x$ is achieved by retaining only the term with index $r_1=0$ in (\ref{pdf3}) along with $R_1=0$ \cite{pks}.  
	\end{IEEEproof}
	
	{\bf{Remark:}} \textit{Note that due to the term $\lambda_4$ in (\ref{8}), $\mathcal{P}_{\textmd{out}}(\gamma_{\textmd{th}})=1$, for $\gamma_{\textmd{th}}\geq\frac{1}{\lambda_1}$. It is a so called SNDR ceiling effect that ceases the communication beyond a threshold $\gamma_{\textmd{th}}=\frac{1}{\lambda_1}$ under RFHI.} \textit{Upon substituting the appropriate parameters in (\ref{poutas}) and taking $\eta_s=\eta_u=\eta$, for $\gamma_{\textmd{th}}<\frac{1}{\lambda_1}$, one can visualize that $\mathcal{P}_{\textmd{out}}(\gamma_{\textmd{th}})\propto\frac{1}{\eta^M}$. Hence, the achievable diversity order of considered HSTN is only $M$ for both Nakagami-\emph{m} and Rician fadings of terrestrial links due to the bottleneck by SR channel.}

	\section{Numerical Results}\label{num}
	\vspace{-0.1cm}
	Here, using \cite{guo}, we set the satellite parameters  as $\mathcal{T}=300$ K, $\mathcal{W}=15$ MHz, $c=3\times10^{8}$m/s, $\mathrm{d}_{u}=35,786$ Km, $f_\mathrm{c}=2$ GHz, $\vartheta_{u}=4.8$ dB, $\vartheta_{s}=53.45$ dB, $\theta_{u}=0.8^{\circ}$, $\theta_{u 3\textmd{dB}}=0.3^{\circ}$, and $(m_{su}, \flat_{su}, \Omega_{su})=(1, 0.063, 0.0007)$ for heavy shadowing.  
	For UAVs, we set the parameters $v_{1,i}\sim[0.1,30]$ m/s, $v_{2,i}\sim[0,40]$ m/s, $H=40$ m, $R=80$ m, $\Omega_{ud}=1$, and $\alpha=2$. Here, $p_\mathrm{s}=0.5$ is set by adjusting the distribution of $T_{\mathrm{s}}$. Furthermore, we set $\eta_s=\eta_u=\eta$ as SNR. 
	
\begin{figure}[t]
		\centering
		\captionsetup{justification=centering}
		\includegraphics[width=4.0in]{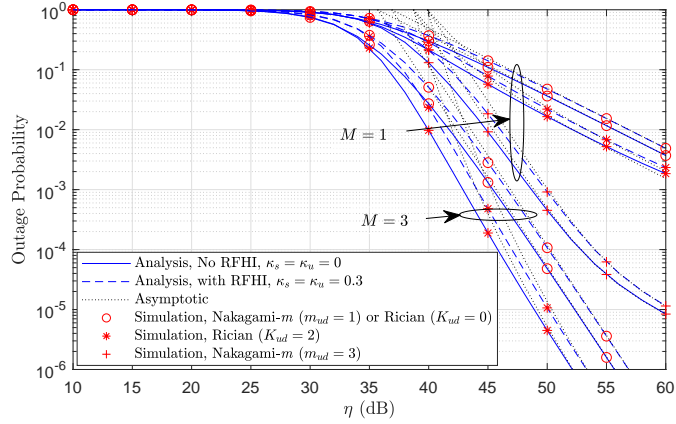}
		\caption{OP versus $\eta$ under terrestrial Nakagami-\emph{m}/Rician fading.}
		\label{fi11}
	\end{figure}

	Fig. \ref{fi11} plots the curves OP versus SNR $\eta$ for considered HSTN. We set  $\gamma_\textmd{th}=0$ dB and $\kappa_s=\kappa_u=0.3$ for RFHI and $\kappa_s=\kappa_u=0$ for perfect hardware. Here, we can see our unified OP and asymptotic OP analyses for both types of terrestrial fading scenarios are in well agreement with the simulations. By observing the curves for $M=1$ with $m_{ud}=1$ and $3$ along with $K_{ud}=0$ and $2$, we conclude that the system diversity order is not affected by the UAV-to-UE links' fading channels. However, when $M$ changes from $1$ to $3$, the slope of various OP curves changes and confirms a diversity order of $M$. We further see that OP of the system deteriorates in the presence of RFHI without affecting the system diversity. 
	
\begin{figure}[!t]
		\centering
		\includegraphics[width=4.0in]{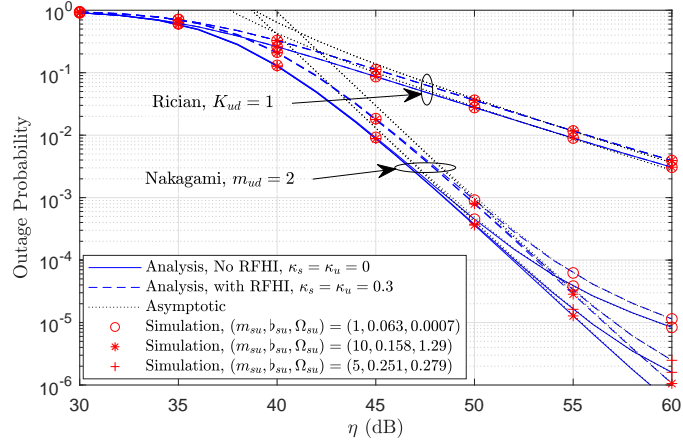}
		\caption{OP versus $\eta$ under various SR fading scenarios.}
		\label{fi13}
	\end{figure}

	Fig. \ref{fi13} plots the curves OP versus SNR for different SR fading scenarios. Here, we set $M=1$ and parameters $(m_{su}, \flat_{su}, \Omega_{su})=(1,0.063,0.0007)$, $(10,0.158,1.29)$, and $(5,0.251,0.279)$ for heavy, light, and average fading scenarios, respectively \cite{mbh}. We observed a performance bottleneck effect under terrestrial Rician links where the OP for all three fading scenarios is indistinguishable. This is due to the dominance of second-hop SNR term as it varies inversely with order unity in asymptotic OP under Rician fading. However, for Nakagami fading, the OP of the system improves when fading changes from heavy to average and then to light shadowing at high SNR. Here, the second-hop SNR term in asymptotic OP varies inversely with order $m_{ud}$. Hence, the first-hop SNR term dominates and impact of SR fading is reflected.
	
\begin{figure}[!t]
		\centering
		\includegraphics[width=4.0in]{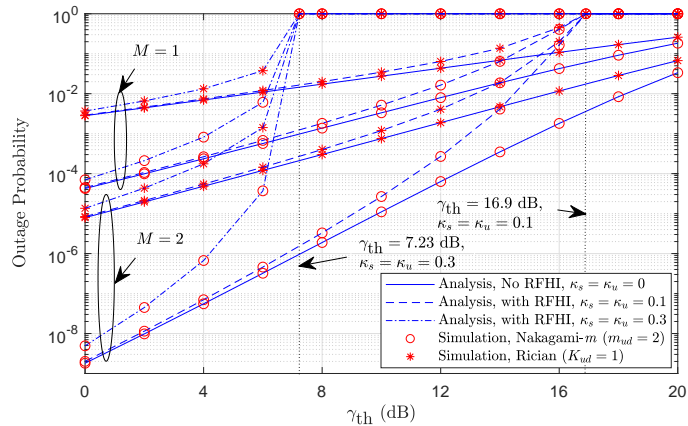}
		\caption{OP versus $\gamma_{\textmd{th}}$ under terrestrial Nakagami-\emph{m}/Rician fading.}
		\label{fi12}
	\end{figure}

	Fig. \ref{fi12} plots the curves OP versus threshold $\gamma_\textmd{th}$ for considered HSTN. We plot curves for two values of RFHI parameters, i.e., $\kappa_s=\kappa_u=0.1$ and $0.3$. Here, we see that for given terrestrial channel fading and RFHI parameters, an SNDR ceiling occurs at certain value of $\gamma_\textmd{th}$ forcing the overall system in outage. The SNDR ceiling threshold is $16.9$ dB and $7.23$ dB for RFHI parameters $0.1$ and $0.3$, respectively. Thus, a higher value of RFHI parameters induces SNDR ceiling at relatively lower thresholds. Moreover, the deviation in OP of system with RFHI from perfect hardware case is relatively severe near ceiling threshold if $M$ increases from $1$ to $2$.

		
		
	\section{Conclusion}\label{con}
	We have analyzed the OP of a multiple AF UAV relay-assisted HSTN in unified form over Nakagami-\emph{m} and Rician terrestrial fading, where each node is corrupted by the RFHI. 
	We examined that RFHI have following effects on the performance of HSTNs: (a) the system diversity order remains unaffected by the RFHI and type of terrestrial channel fading; (b) system is forced into outage at certain threshold which depends only on RFHI parameters.

\end{document}